\documentclass[aps,prr,twocolumn,floats,final,superscriptaddress,citeautoscript,longbibliography]{revtex4-2}

\usepackage{amssymb}
\usepackage{graphicx}
\usepackage{amsmath}
\usepackage{amsthm}
\usepackage{braket}
\usepackage{scrextend}
\usepackage{footmisc}
\usepackage[colorlinks=true,linkcolor=blue,citecolor=blue,urlcolor=blue]{hyperref}
\usepackage[svgnames]{xcolor}
\usepackage{tikz}
\usepackage{comment}

\usepackage{times}
\usepackage{lineno}

\def\brho{{\boldsymbol \brho}}

\def\bp{{\boldsymbol p}}

\def\br{\boldsymbol{r}}
\def\bj{{\boldsymbol j}}
\def\bd{{\boldsymbol d}}
\def\bv{{\bf v}}

\def\bz{{\boldsymbol z}}

\def\bR{{\boldsymbol R}}

\def\bP{{\boldsymbol P}}
\def\bA{{\boldsymbol A}}

\def\calL{\mathcal{L}}
\def\calN{\mathcal{N}}
\def\calM{\mathcal{M}}

\def\grad{\mbox{\boldmath $\nabla$}}

\def\pa{\partial}
\def\nn{\nonumber}

\def\ra{\rangle}



\begin{document}

\title{Bosonic quantum Hall droplets in rapidly rotating  two-dimensional Bose-Einstein condensates}

\author{Zhen Cao}
\altaffiliation{These authors contributed equally to this work.}
\author{Siying Li}
\altaffiliation{These authors contributed equally to this work.}
\author{Zhendong Li}
\author{Xinyi Liu}
\affiliation{State Key Laboratory of Information Photonics and Optical Communications, School of Physical Science and Technology, Beijing University of Posts and Telecommunications, Beijing 100876, China
}

\author{Zhigang Wu}
\email{wuzhigang@quantumsc.cn}
\affiliation{Quantum Science Center of Guangdong-Hong Kong-Macao Greater Bay Area (Guangdong), Shenzhen 508045, China
}
\author{Mingyuan Sun}
\email{mingyuansun@bupt.edu.cn}
\affiliation{State Key Laboratory of Information Photonics and Optical Communications, School of Physical Science and Technology, Beijing University of Posts and Telecommunications, Beijing 100876, China
}

\date{\today}

\begin{abstract}
Recent experiments demonstrate that rapidly rotating Bose-Einstein condensates (BECs) near the lowest Landau level can self-organize into interaction-driven persistent quantum Hall droplet arrays. Inspired by this discovery, we investigate the formation and dynamics of single quantum Hall droplet and droplet arrays in rapidly rotating BECs. Guided by a rigorous theorem on localized many-body states for 2D interacting systems in a magnetic field, we construct single quantum Hall droplet and droplet array states which are shown to be stationary solutions to the Gross-Pitaevskii equation in the rotating frame. The single quantum Hall droplet is shown to be dynamically stable, which underpins its role as the basic unit in a droplet array. The stability of the quantum Hall droplet arrays is demonstrated by their dynamic formation from a phase engineered initial condensate.  Our study sheds light onto the nature of the quantum Hall droplet state in a rapidly rotating BEC and offers a new approach for generating and manipulating quantum Hall droplet arrays through designing the initial condensate phase.

\end{abstract}

\maketitle

\emph {Introduction.---}Rotating Bose-Einstein condensates (BECs) have been subjects of intense interest even since the first realizations of the atomic BEC~\cite{Davis1995,Anderson1995}.  Earlier focus has been on the creation of quantized vortices in these systems since they are one of the most definitive signatures of superfluidity~\cite{Matthews1999,Madison2000,Haljan2001,Abo-Shaeer2002,KASAMASTU2009}. As experimental control of these systems progresses,  they have emerged as a crucial platform for studying topological quantum phases and quantum fluid dynamics~\cite{Bloch2008.A,Galitski2013.A,Senthil2013.A,
Goldman2014.A,Aidelsburger2014.A,
Baggaley2018,Gra2018.A,Chalopin2020.A,Richaud2020,Richaud2023,Dutta2023.Z,Gu2023.Z,Oktel2023,White2024.Z}. In particular, under the condition of rapid rotation, the atomic gases in the rotating frame can simulate those in a strong magnetic field, thus allowing for the study of quantum Hall physics in bosonic systems~\cite{Ho2001}. Indeed, various theories have predicted that as the filling fraction decreases the rotating BEC may transition from the vortex lattice phase in the mean-field quantum Hall regime to various strongly correlated phases in the fractional quantum Hall regime~\cite{Cooper2001,Sinova2002,Baym2004,Wu2007,Cooper2008.B,Fetter2009.B}. 

A critical component of reaching the quantum Hall regimes experimentally is the ability to prepare the atoms in the lowest Landau level (LLL),  where the kinetic energy is quenched and the interactions dominate. Although signatures of LLL physics have been previously observed by spinning the BEC up close to the centrifugal limit~\cite{Schweikhard2004,Bretin2004}, recent experiments have developed a powerful geometric squeezing method that can prepare the rotating BEC in a single Landau gauge wave function in the LLL~\cite{Fletcher2021,Mukherjee2022,Crpel2024}. Moreover, these later experiments were able to achieve filling fractions that are an order of magnitude lower compared to those in earlier ones, placing the system firmly in the mean-field quantum Hall regime.  Interestingly, this Landau gauge BEC does not evolve into a vortex lattice but instead spontaneously form 1D persistent droplet arrays~\cite{Mukherjee2022}, characterized by density modulations at the magnetic length scale and vortex streets between the individual droplets. We refer to these droplet states as bosonic quantum Hall droplets which  are analogues of those in electronic liquids under strong magnetic fields. Although the crystallization instability of the Landau gauge BEC has been explained by the condensation of magneto-rotons,  the underlying mechanism for the bosonic quantum Hall droplet formation and its inherent stability are still open questions.

In this work, we develop a theoretical framework to understand the nature of the bosonic quantum Hall droplet states in rapidly rotating BECs, by combining rigorous analytic insights with extensive numerical simulations of the Gross-Pitaevskii equation (GPE). We first prove a theorem for 2D interacting systems in a uniform magnetic field, showing that they host many-body eigenstates that can localize around any position in the 2D plane. In the context of rapidly rotating BEC, we use this theorem to construct the bosonic quantum Hall droplet states. Although these states are reminiscent of the quantum droplet states in bosonic mixtures and dipolar gases~\cite{Petrov2015,Petrov2016,Luo2020,Simos1,Simos2,Simos3,Simos4},  their formation is due to the presence of strong gauge fields rather than the competition of mean-field and zero-point energies. The linear stability of the quantum Hall droplet is explicitly demonstrated by solving the Bogoliubov-de Gennes equations in the rotating frame.  Furthermore, we devise a method to construct stationary quantum Hall droplet arrays of any configuration using the single droplet state as the unit. To show that such arrays are stable structures, we adopt a phase-engineering approach and achieve  controlled  dynamic formation of 2D droplet lattice and circular droplet array.  Finally, we show that  such an approach can even create droplet arrays in which the quantum Hall droplets perform cyclotron motion. Our work places the concept of  bosonic quantum Hall droplet state in a rotating BEC on a firm theoretical ground and establishes a novel approach for generating and manipulating droplet arrays, which may provide a pathway to exploring correlated quantum dynamics in these systems.

\emph {Localized eigenstates theorem.---}
We begin by proving a theorem for 2D interacting systems in a magnetic field which will form the basis of our discussion on quantum Hall droplet states in such systems. We show that in these systems i) there exists a class of localized many-body eigenstates that are related to each other by the magnetic translation of the center of mass degree of freedom and ii) these localized states can be set in motion as a whole and form a class of dynamical many-body states for which the probability density performs cyclotron motion without change. The Hamiltonian of a 2D interacting system in a magnetic field reads
\begin{align}
\hat H = \sum_j  \frac{1}{2m} \left [\hat \bp_j - \bA(\hat\br_j)\right ]^2   + \sum_{j<k} V(\hat \br_j - \hat \br_k),
\end{align}
where $\bA(\br) = m\Omega\hat z \times \br$ is the vector potential of a uniform magnetic field $\boldsymbol B = 2m\Omega\hat z$ and  $V(\br)$ is the pairwise particle interaction.  This Hamiltonian is equivalent to that of a harmonically trapped system in the rotating frame in the rapid rotation limit, i.e.,  $\hat H = \hat H' - \Omega \hat L_z$, where $\hat H' =  \sum_j  (\hat \bp_j^2/2m +m\Omega^2\hat \br_j^2/2)   + \sum_{j<k} V(\hat \br_j - \hat \br_k)$ is the Hamiltonian for the trapped system, $\Omega$ is both the angular velocity of rotation and the trap frequency and $\hat L_z$ is the total angular momentum. Due to the rotational invariance, an eigenstate $|\Psi_\alpha\ra$ of this trapped system $\hat H'$  is also an eigenstate of $\hat H$ with energy $E_\alpha$. Such a state is naturally localized around the origin (i.e., the center of the trap) as the expectation value of the center of mass coordinate is zero. Let's consider the dynamical evolution of the system when prepared at $t=0$ in the state 
\begin{align}
|\Psi(0)\ra \equiv e^{i\left (\bp_0\cdot \hat \bR - \br_0\cdot \hat \bP\right )} |\Psi_\alpha\ra,
\end{align}
where $\hat \bR = \frac{1}{N} \sum_j \hat \br_j
$  is the center of mass operator and $\hat \bP = \sum_j \hat \bp_j$ is  the total canonical momentum operator. The dynamical state at time $t$ is ($\hbar = 1$ throughout the paper)
\begin{align}
|\Psi(t)\ra = e^{-iE_\alpha t}e^{i\left (\bp_0\cdot \hat \bR(-t) - \br_0\cdot \hat \bP(-t)\right )} |\Psi_\alpha\ra,
\label{Psit}
\end{align}
where $\hat \bR(t)  =e^{i\hat H t}  \hat \bR e^{-i\hat H t} $ and  $\hat \bP(t) =e^{i\hat H t}  \hat \bP e^{-i\hat H t} $ are Heisenberg operators. Using the Heisenberg equation of motion, we obtain
\begin{align}
\label{Hs1}
\frac{d \hat \bR}{dt} &= \frac{1}{M}\hat \bP - \Omega \hat \bz \times \hat \bR \\
\frac{d\hat \bP}{dt}& = -M\Omega^2 \hat \bR - \Omega \hat \bz \times \hat\bP,
\label{Hs1p}
\end{align}
where $M = N m$ and $\hat \bz$ is the unit vector along the $z$-direction. Here it is important to note that in deriving Eq.~(\ref{Hs1p}) we have used the fact that the total momentum $\hat \bP$ commutes with any pairwise interaction, i.e.,  $[\hat \bP(t),  \sum_{j<k} V(\hat \br_j - \hat \br_k)] = 0$.  By further using the Heisenberg equation of motion, we arrive at
\begin{equation}
\ddot{\hat \bR}  = -2\Omega \hat\bz \times \dot{\hat \bR}
\end{equation}
\begin{equation}
\dddot{\hat\bR} = -4\Omega^2 \dot{\hat \bR}.
\label{Hs2}
\end{equation}
The above equations can be solved and one can obtain~\cite{SM} 
\begin{align}
\bp_0\cdot \hat \bR(-t) - \br_0\cdot \hat \bP(-t) = \bp_0(t)\cdot \hat \bR - \br_0(t)\cdot \hat \bP
\label{BH}
\end{align}
such that 
\begin{align}
|\Psi(t)\ra =  e^{-iE_\alpha t}e^{i\left (\bp_0(t)\cdot \hat \bR - \br_0(t)\cdot \hat \bP \right )}|\Psi_\alpha\ra.
\label{Psit2}
\end{align}
 Here $\br_0(t) = \br_0 - \bd(0) +\bd(t)$ and $
\bp_0(t) = M\dot{\br}_0(t) + M\Omega \hat z \times \br_0(t)$, where $\bd(t) = (d_x(t),d_y(t))$ is given by
\begin{align}
 \begin{pmatrix}
 d_x(t) \\ d_y(t)
 \end{pmatrix} =  \begin{pmatrix}
 \cos\omega_c t & \sin\omega_c t \\
 -\sin\omega_c t& \cos\omega_c t
 \end{pmatrix}  \begin{pmatrix}
 d_x(0) \\d_y(0)
 \end{pmatrix},
 \end{align} 
 with $\bd (0) = \frac{1}{2}\br_0 + \frac{1}{2M\Omega}\hat z\times\bp_0$ and $\omega_c = 2\Omega$. 
 
In general, $\br_0(t)$ and $\bp_0(t)$ are time-dependent functions except when $\br _0=- \frac{1}{M\Omega}\hat z \times \bp_0$ for which $\bd(0) = 0$. In this case, both $\br_0(t) = \br_0$ and $\bp_0(t) = \bp_0$ are time-independent and we obtain the following eigenstates from Eq.~(\ref{Psit2})
  \begin{align}
|\Psi\ra = e^{-i\left ( \hat \bP + M\Omega \hat z\times \hat\bR \right )\cdot \br_0}|\Psi_\alpha\ra.
\end{align} Each  of these states shares the same energy $E_\alpha$ and is related to the original localized state $|\Psi_\alpha\ra$ by a magnetic translation of the center of mass. More explicitly, the many-body wave function of such a state is 
\begin{align}
\Psi(\br_1,\cdots,\br_N) &= e^{i  \sum_j \bA(\br_0)\cdot \br_j}  \nn \\
&\times\Psi_\alpha(\br_1-\br_0,\cdots,\br_N-\br_0),
\label{Psir}
\end{align}  
where an irrelevant constant phase factor is dropped.
 When $\br _0 \neq -\frac{1}{M\Omega}\hat z \times \bp_0$, it is clear that  $\br_0(t)$ and $\bp_0(t)$, respectively, represent the trajectory and the canonical momentum of a particle of mass $M = Nm$ performing the cyclotron motion of frequency $\omega_c = 2\Omega$, whose orbit is centered at $\br_0$ and has radius given by $\bd(0)$.  In this case, 
 the many-body wave function can be written as
\begin{align}
\Psi(\br_1,\cdots,\br_N;t) &= e^{-iE_\alpha t}e^{i\sum_j [m{\dot \br}_0(t)  +\bA(\br_0(t))]\cdot \br_j  }  \nn \\
&\times\Psi_\alpha(\br_1-\br_0(t),\cdots,\br_N-\br_0(t)).
\label{Psirt}
\end{align}
This means that for $\bd(0) \neq 0$ the probability density of this state $|\Psi(\br_1,\cdots,\br_N;t) |^2 = |\Psi_\alpha(\br_1-\br_0(t),\cdots,\br_N-\br_0(t))|^2$ simply performs cyclotron motion rigidly following the trajectory given by $\br_0(t)$.  Physically our theorem reflects the fact that 2D systems in a magnetic field are similar to those confined in a harmonic potential in that the center of mass degree of freedom is decoupled from the internal ones even in the presence of interaction~\cite{Dobson1994,BialynickiBirula2002,Wu2011,Wu2014}. We emphasize that our derivation is mathematically rigorous and valid for any choice of the pairwise interaction. Therefore, the theorem is universal in the sense that it does not depend on particle statistics nor on any specific form of the pairwise interaction. It would be interesting to generalize this theorem to higher-dimensional systems and we will leave this to future study. In the following we will make use of our theorem to construct various quantum Hall droplet states for a rapidly rotating BEC. 

\emph {Single quantum Hall droplet state.---}In cold atomic experiments, quasi-2D confinement is ensured by making the axial trap frequency $\omega_z$ much larger than the radial frequency $ \omega_\perp$, such that the effective 2D interaction strength is  $g=a_s\sqrt{{8\pi \omega_z}/{m}}$ where $a_s$ is the s-wave scattering length. When the thermal and interaction energies are significantly smaller than $\hbar\omega_z$, the transverse degree of freedom can be effectively integrated out, resulting in an effective two-dimensional system. In the rapid rotation limit $\Omega = \omega_\perp $, the quasi-2D Bose condensate in the rotating frame is described by the 2D Gross-Pitaevskii equation~\cite{Gross1961,Pitaevskii1961,Gross1963}
\begin{align}
i \frac{\pa}{\pa t}\psi = \left [   -\frac{1}{2m} (\nabla -i\bA(\br)) ^2  + g |\psi|^2 \right ] \psi,
\label{TDGProtate}
\end{align}
where the condensate wave function $\psi$ is normalized to the number of atoms $N$. 
\begin{figure}[t]
\centering
\includegraphics[width=0.48\textwidth]{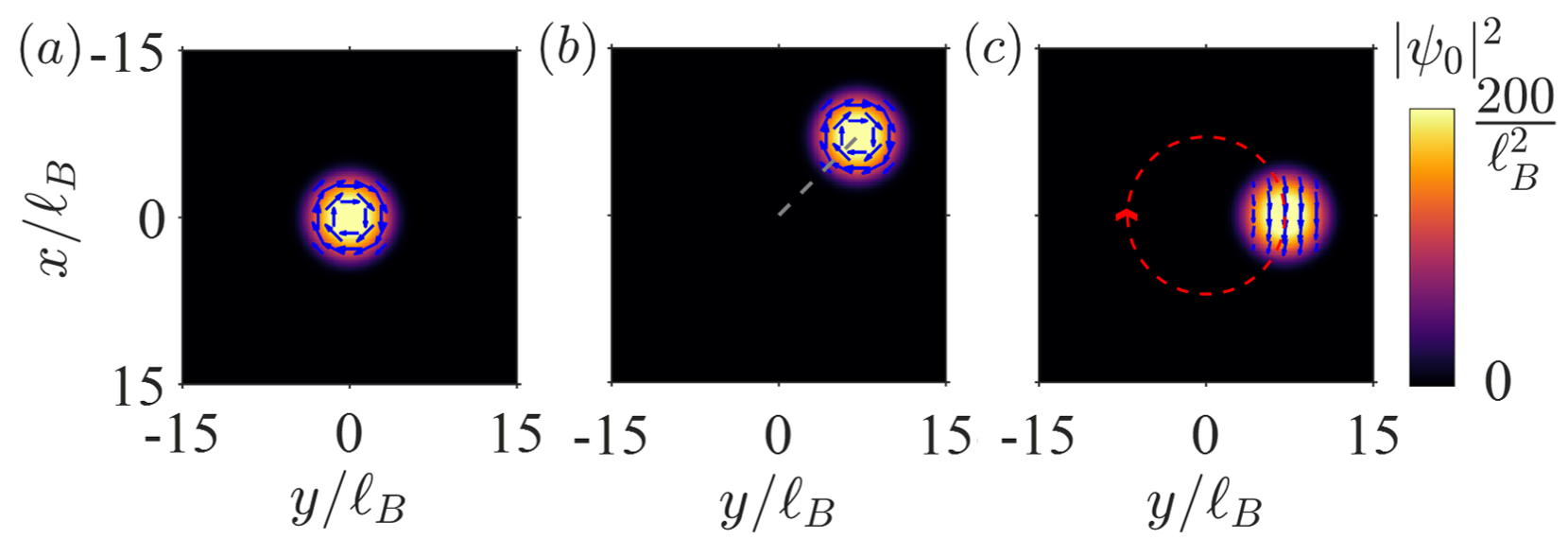}
\caption{\label{fig:FIG1} A quantum Hall droplet centered at the origin (a), at $\br_0 = (-7,7)\ell_B$ (b) and performing the cyclotron motion as a whole (c). The arrows denote the current field in the droplets. Here $N =8\times10^3 $, $g = 1.25\times10^{-2}\omega_c\ell_B^2$, $\omega_c = 2\pi \times177.2$ Hz and $\ell_B \approx 1.6\mu$m. The last three are common parameters used in all calculations.}
\end{figure}
 In the first instance, we may look for droplet states from stationary solutions to the above equation. There are two types of stationary solutions that can be easily verified. 
 The first is $\psi(\br,t) =  \psi_0(\br)e^{-i\mu t}$, where $ \psi_0(\br)$ is the ground state of the static trapped condensate satisfying 
\begin{align}
\left ( -\nabla^2/2m+ m\Omega^2 r^2/2 + g |\psi_0|^2\right ) \psi_0(r) = \mu \psi_0(r),
\end{align} 
and $\mu$ is the chemical potential. 
The second is $\psi(\br,t) = \tilde\psi_0(x)e^{-im\Omega xy}e^{-i\tilde \mu t}$~\cite{Pitaevskii2016}, where $\tilde \psi_0(x)$ is given by
 \begin{align}
\left ( -\pa_x^2/2m+ 2m\Omega^2 x^2 + g|\tilde \psi_0|^2\right ) \tilde \psi_0(x) = \tilde \mu \tilde \psi_0(x).
\label{tildepsi}
\end{align} 
We focus on the former because the second type is known to be dynamically unstable~\cite{Mukherjee2022}. 
From Eq.~(\ref{Psir}) and the relation between the condensate wave function and the many-body wave function, we  find that the state
\begin{align}
\psi_0(\br;\br_0) \equiv e^{i\bA(\br_0)\cdot \br} \psi_0(\br - \br_0)
\label{phir}
  \end{align} 
must also be a stationary solution to Eq.~(\ref{TDGProtate}). Aside from these stationary solutions, Eq.~(\ref{Psirt}) suggests that 
Eq.~(\ref{TDGProtate}) must accept dynamical states $\psi(\br,t) = \psi_0(\br;\br_0(t)) e^{-i\mu t}$ as solutions, where 
  \begin{align}
\psi_0(\br;\br_0(t))  \equiv  e^{i [m{\dot \br}_0(t)  +\bA(\br_0(t))]\cdot \br  } \psi_0(\br- \br_0(t))
\label{phirt}
  \end{align}
describes a condensate performing the cyclotron motion $\br_0(t)$ as a whole.  Examples for  $\psi_0(\br)$ and the related two states  $\psi_0(\br;\br_0)$ and $\psi_0(\br;\br_0(t))$ are shown in Fig.~\ref{fig:FIG1}. Naturally, the stationary and the moving droplets can be distinguished by the current field $\bj(\br,t) =m |\psi|^2 \boldsymbol v(\br,t) $, where 
\begin{align}
 \boldsymbol v(\br,t)= \grad \varphi (\br,t)- \bA (\br)
\end{align}
 is the velocity field. Here $\varphi(\br,t)$ is the phase of the condensate wave function $\psi(\br,t)= |\psi(\br,t)|e^{i\varphi(\br,t)}$.

\emph {Dynamical stability.---}For the states in Eq.~(\ref{phir})  and Eq.~(\ref{phirt}) to be genuine quantum Hall droplet states, they must be dynamically stable. Since they are internally the same state as $\psi_0(\br)$, it is sufficient to examine the dynamical stability of the latter. For this purpose, we calculate the quasi-particle spectrum in the rotating frame using the Bogoliubov-de Gennes (BdG) equation, which can be derived by substituting $\psi(\br,t)  = \psi_0(\br)e^{-i\mu t } + \delta \psi(\br,t)$ in Eq.~(\ref{TDGProtate}), where 
\begin{align}
 \delta \psi(\br,t)= e^{-i\mu t}\sum_{nl} \left [ u_{nl} (\br) e^{-i\epsilon_{nl} t} - v^*_{nl}(\br) e^{i\epsilon_{nl} t} \right ].
\end{align}
The Bogoliubov amplitudes $u_{nl} (\br),v_{nl}(\br)$ and the excitation energy $\epsilon_{nl}$ are labeled by the principal quantum number $n$ and the angular momentum quantum number $l$. The amplitudes can be further expanded in terms of the Landau level wave functions $\phi_{n'l}(\br)$ as $u_{nl} (\br) = \sum_{n'}U_{nn'}\phi_{n'l}(\br)$ and  
$v_{nl} (\br) = \sum_{n'  }V_{nn'}\phi_{n'l}(\br)$. In symmetric gauge, the $n'$-th Landau level wave function is
 \begin{align}
\phi_{n'l}(r,\theta) = \calN
\left( \frac{r}{\sqrt{2} \ell_B} \right)^{|l|} 
L_{n'}^{(|l|)}\left( \frac{r^2}{2\ell_B^2} \right) 
e^{i l \theta} 
e^{- \frac{r^2}{4 \ell_B^2}},
\end{align}
where $\calN  = \sqrt{\frac{n'!}{2\pi \ell_B^2  (n' +| l|)!}} $ is the normalization constant, $\ell_B = \sqrt{1/(2m\Omega)}$ is the magnetic length and $L_n^{(l)}(x)$ is the generalized Laguerre polynomials 
$
L_n^{(l)}(x) = \frac{1}{n!} x^{-l} e^x \frac{d^n}{dx^n} \left( e^{-x} x^{n+l} \right)
$.
As the angular momentum is a good quantum number, the BdG equation takes the form of 
\begin{align}
\label{BdGM1}
\sum_{n'} \left [ \calL_{n'' n'}{(l)} U_{nn'} - \calM_{n'' n'} {(l)} V_{nn'}\right ]&= \epsilon_{nl} U_{nn''}; \\
\sum_{n'} \left [  \calM_{n'' n'} {(-l)} U_{nn'}- \calL_{n'' n'}{(-l)} V_{nn'} \right ]&= \epsilon_{nl} V_{nn''},
\label{BdGM2}
\end{align}
where 
$
\calM_{n'n} {(l)} = g \int  \psi^2_0(\br)  \phi^*_{n'l}(\br)\phi_{nl}(\br) d\br$ and 
$\calL_{n'n} {(l)} =[ (2n'+1+|l|-l)\Omega - \mu ]\delta_{n'n} +2\calM_{n'n} {(l)}$. 

\begin{figure}[t]
\centering
\includegraphics[width=0.48\textwidth]{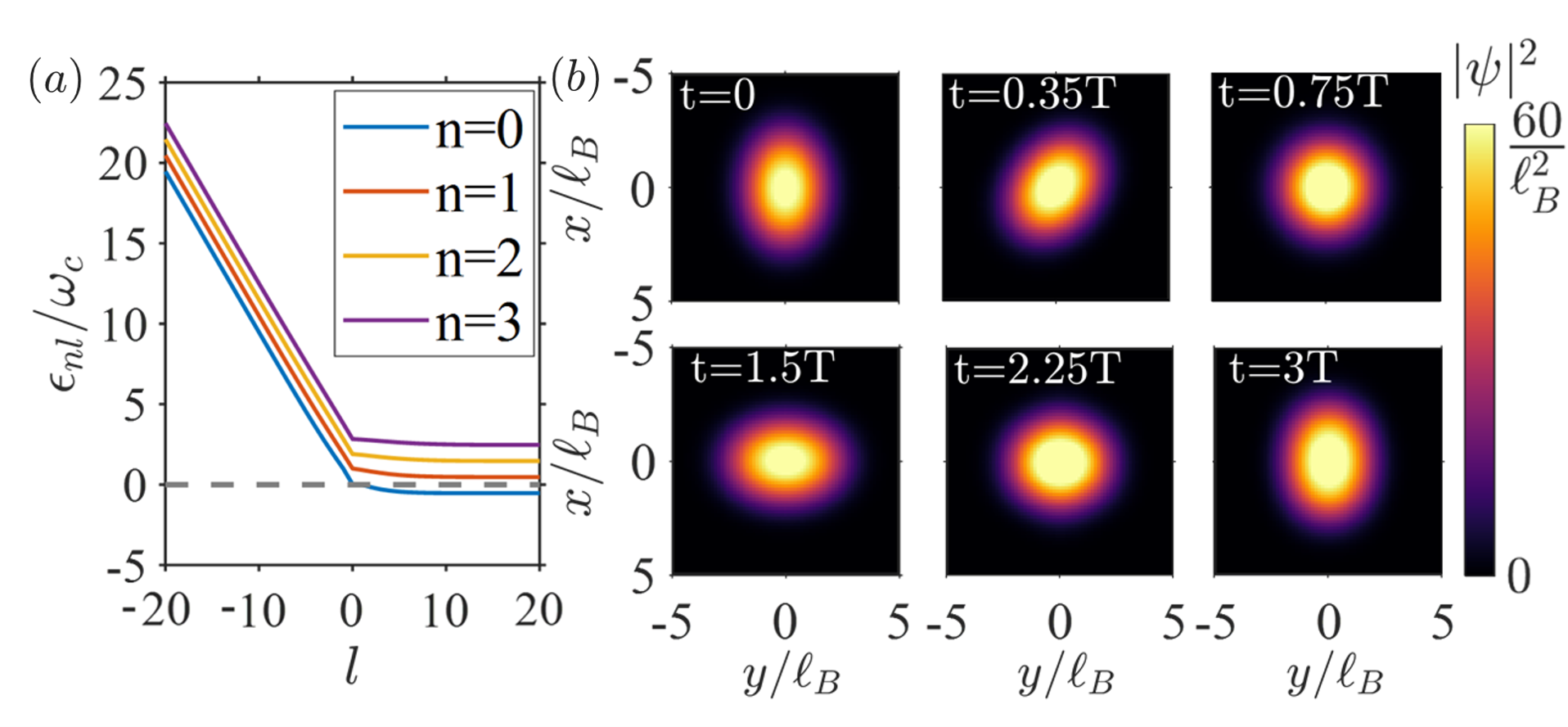}
\caption{\label{fig:FIG2} (a) Quasi-particle spectrum of a single quantum Hall droplet state. (b) Dynamic evolution of a quantum Hall droplet perturbed from equilibrium. Here $T = 2\pi/\Omega$ and $N = 800$ . }
\end{figure}

Crucially, we find that the excitation spectra obtained by solving Eqs.~(\ref{BdGM1})-(\ref{BdGM2}) are all real (see Fig.~\ref{fig:FIG2}(a)), ensuring that the state $\psi_0(\br)$ is dynamically stable. This is in stark contrast with the excitation spectra of the second type of stationary state $\tilde\psi_0(x) e^{-im\Omega xy}$ in Eq.~(\ref{tildepsi}) which always contains an imaginary branch underlying the exponential instability of the state. The dynamical stability of the quantum Hall droplet is also directly verified by perturbing the condensate from equilibrium [see Fig.~\ref{fig:FIG2}(b)]. For example, when slightly squeezed along one direction, the droplet performs oscillations that are well accounted for by excitations determined by the BdG equation~\cite{SM}.  Remarkably, even under strong perturbation for which the linear stability analysis is no longer applicable, the droplet is still stable against further dissociating into smaller subsystems or vortex states~\cite{SM}. Such linear and non-linear stability are essential characteristics of the quantum Hall droplet state. 

There is another salient feature to the droplet in any dynamic state $\psi(\br,t)$ due to the dynamical SO(2,1) conformal symmetry in this system~\cite{SaintJalm2019}. That is both the moment of inertia $  I=\int d\br\psi^* r^2\psi
$ and the physical angular momentum $
\boldsymbol{L}_{\rm p}  =\int d\br \psi^* (\boldsymbol{r}\times \boldsymbol{v}) \psi
$ of the droplet  exhibit an undamped oscillation with frequency $2\Omega$. It can be shown that~\cite{Kagan1996,Pitaevskii1997,SaintJalm2019}
\begin{align}
\label{equ:I2}
  I(t)&=({2}/{m\Omega^2})(E_{\rm tot}+\Delta E \cos2\omega t); \\
\boldsymbol{L}_{\rm p} (t) &=\boldsymbol{L}_{\rm c} -I(t), 
\label{equ:LR2}
\end{align}
where $E_{\rm tot}$ is the total energy of the system, $E_{\rm pot}(t)=m\Omega^2 \int d\br  r^2\left| \psi (\br,t) \right|^2/2$ is the trapping potential energy, $\Delta E=2E_{\rm pot}(0)-E_{\rm tot}$ and $ \boldsymbol{L}_{\rm c} =\int d\br\psi^* (\boldsymbol{r}\times \boldsymbol{\nabla} \varphi) \psi$ is the canonical angular momentum. The canonical angular momentum, just like the total energy, is conserved in a dynamical state, i.e., $\boldsymbol{L}_{\rm c}(t) \equiv\boldsymbol{L}_{\rm c}(0)$~\cite{Fetter2001,Bao2006_2,Bao2013}. Equations (\ref{equ:I2}) and (\ref{equ:LR2}) will later be used to analyze single quantum Hall droplet behavior in formations of dynamic droplet arrays.   

\begin{figure}[t]
\centering
\includegraphics[width=0.48\textwidth]{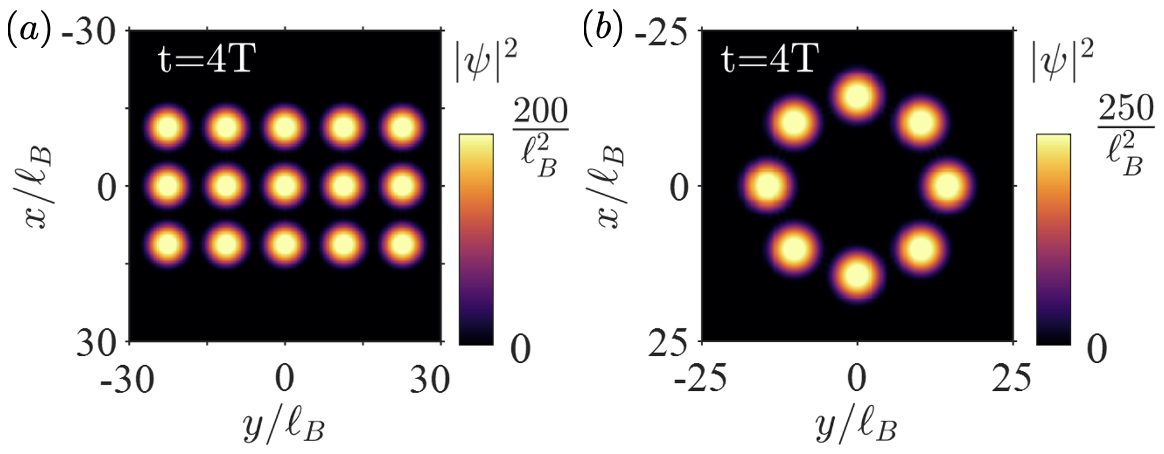}
\caption{\label{fig:FIG3} Stationary droplet arrays. Densities of the droplet arrays at $t = 4T$ are shown, which are indistinguishable from those at $t=0$. The parameters here are $N = 10^5$ (a) and $N = 8\times 10^4$ (b). }
\end{figure}

\emph {Stationary and rotating quantum Hall droplet arrays.---}We now show that the single quantum Hall droplet state in Eq.~(\ref{phir}) can be used in constructing stationary quantum Hall droplet arrays of any configuration. Let's consider a coherent superposition of a number of single quantum Hall droplet states $\psi_j$, each containing $N_j$ atoms and centered at $\br_j$, i.e., 
  \begin{align}
\psi(\br,t) = \sum_{j} e^{i\bA(\br_j)\cdot \br} \psi_{j}(\br - \br_j) e^{-i\mu_j t},
\label{phirtarray}
  \end{align}
  where $\mu_j$ is the chemical potential of the $j$-th quantum Hall droplet. Since each droplet is characterized by a finite radius, we can choose ${\br_j}$ to be sufficiently apart from each other such that there is no overlap between any two droplets, i.e., $\psi_i(\br - \br_i) \psi_j(\br - \br_j) =  0$ for any $i\neq j$. In this case, it can be easily verified $\psi(\br,t)$ in Eq.~(\ref{phirtarray}) is a solution to Eq.~(\ref{TDGProtate}). Since there is no overlap between different droplets, the total density of the condensate is stationary, i.e., 
  \begin{align}
  \rho(\br) = \sum_j  |\psi_j(\br - \br_j)|^2.
  \end{align}
  
\begin{figure}[t]
\centering
\includegraphics[width=0.45\textwidth]{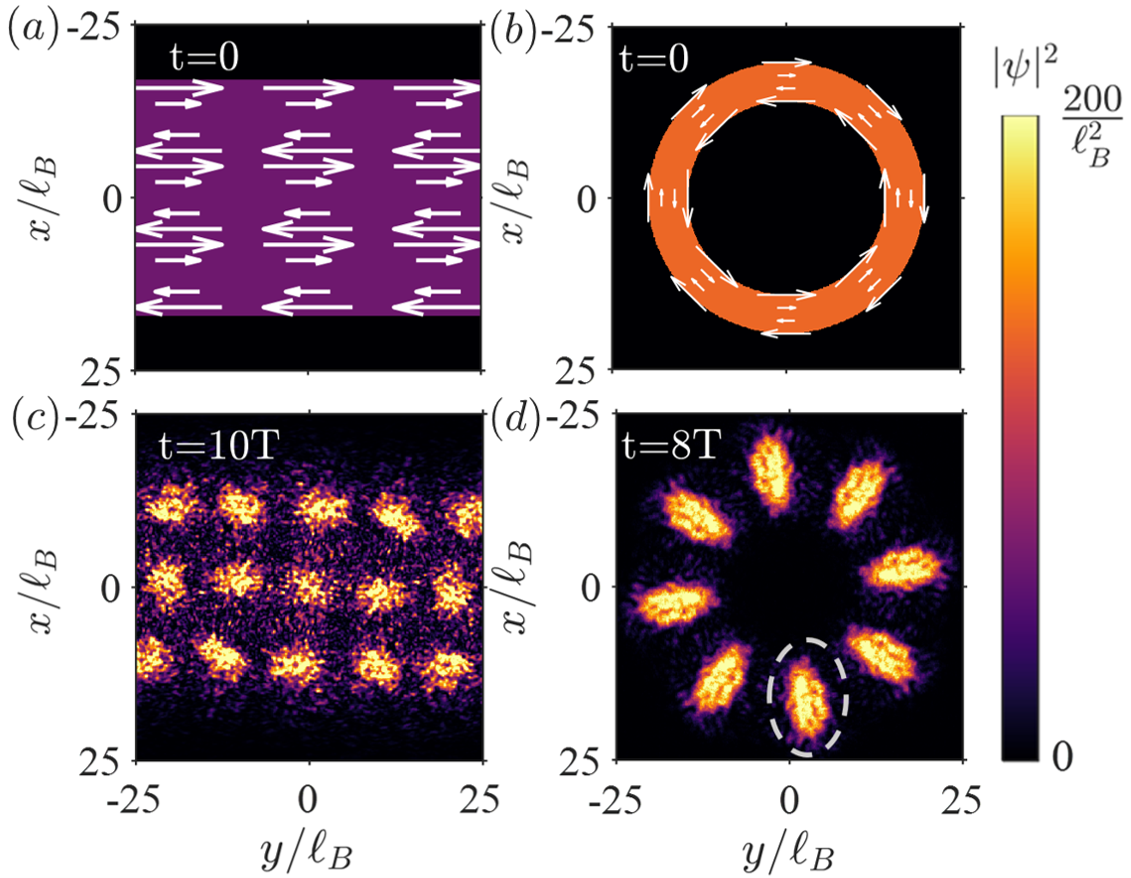}
\caption{\label{fig:FIG4} Formation of dynamic quantum Hall droplet arrays. A  uniform condensate (a)  with phase designed according to Eq.~(\ref{philattice}) evolves into a quantum Hall droplet lattice (c).  Here $N = 2.4\times10^5$ and the dimension  of the initial condensate is $24\sqrt{2}\ell_B\times 80\sqrt{2}\ell_B$. A ring-shaped uniform condensate (b)  with phase designed according to Eq.~(\ref{phiring}) evolves into a circular quantum Hall droplet array (d). Here $N = 8\times10^4$ and the  outer radius and the width of initial condensate are  $14\sqrt{2}\ell_B$ and $4\sqrt{2}\ell_B$ respectively.   }
\end{figure}

We emphasize that such a quantum Hall droplet array can be of any configuration as long as different droplets have no overlaps. As two simple examples, we show in Fig.~\ref{fig:FIG3} a 2D quantum Hall droplet lattice and a circular quantum Hall droplet array constructed using single quantum Hall droplet states containing the same number of atoms. By time evolving these states we have verified that the densities of the arrays so constructed are indeed stationary~\cite{SM}. Similar to our previous discussions of the moving single quantum Hall droplet we can also construct quantum Hall droplet arrays that performs rigid-body rotations at the cyclotron frequency $\omega_c$ by taking
    \begin{align}
\psi= \sum_{j} e^{i [m\dot\br_j(t) + \bA(\br_j(t))]\cdot \br} \psi_{j}(\br - \br_j(t)) e^{-i\mu_j t},
\label{phirtarrayr}
  \end{align}
where $\br_j(t)$ are cyclotron trajectories all centered at the origin. 

Lastly, an interesting question is whether these conclusions still hold if there are overlaps between the droplets in the initial array. If the overlaps are very small our simulations show that the droplet arrays are still stable for a considerable time. A thorough investigation of cases where the overlaps are significant will be left to future studies.

\emph {Formation of dynamic quantum Hall droplet arrays.---}
After establishing  various quantum Hall droplet array solutions to the time-dependent GP equation of the rapidly rotating BEC, a natural question concerns the stability of these states.  This is a considerably more complex problem than the stability of a single quantum Hall droplet because under external perturbations the interaction between adjacent droplets will come into play. Rather than calculating the excitation spectrum, here we demonstrate that droplet arrays can emerge spontaneously out of initial states that are significantly deviated from the stationary droplet array states. The droplet arrays so obtained are necessarily dynamic and can be viewed as the excited states of the stationary droplet array; their formation is alternative evidence that droplet arrays are indeed stable structures of the rapidly rotating BEC.  

\begin{figure}[b]
\centering
\includegraphics[width=0.48\textwidth]{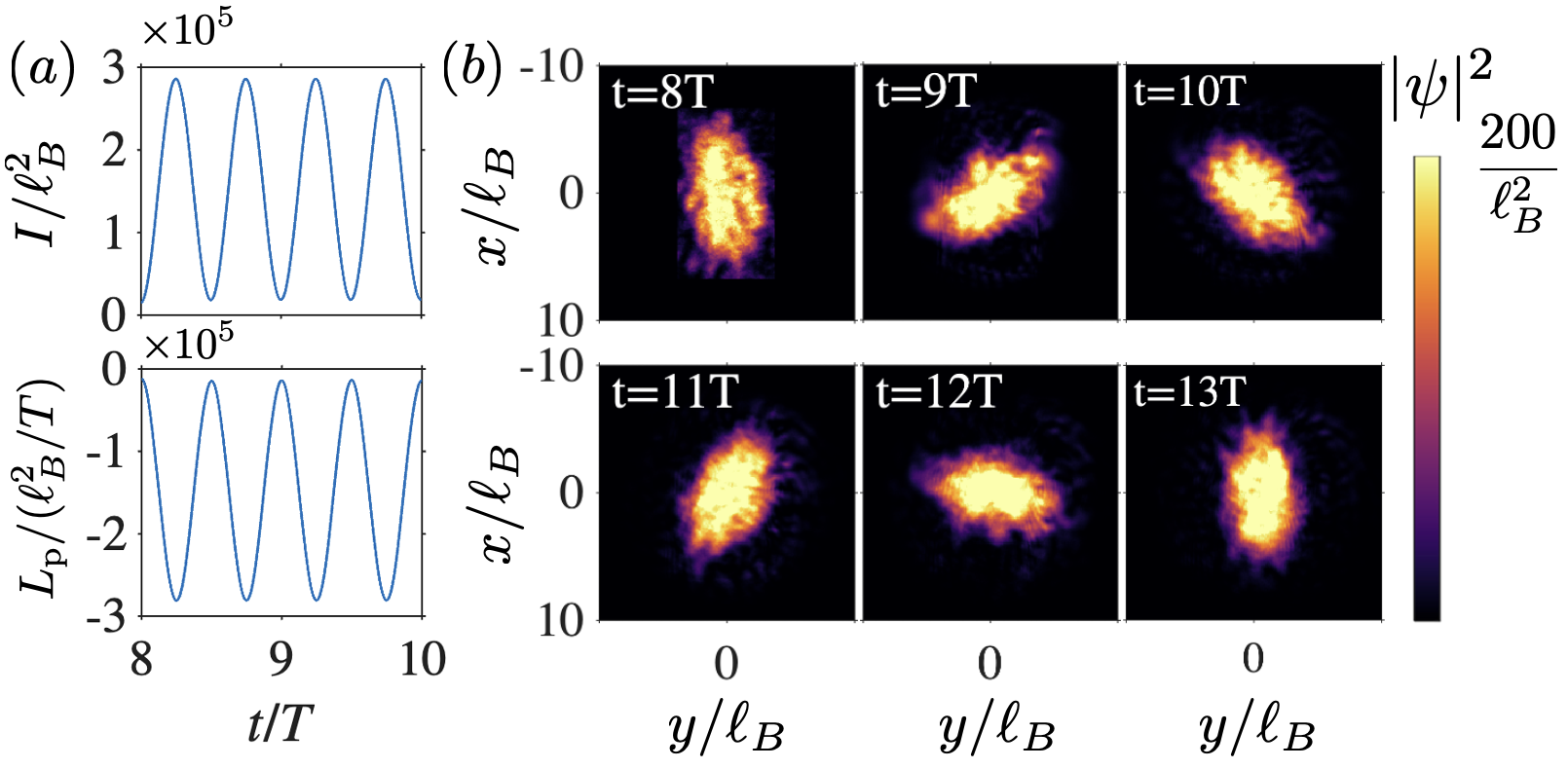}
\caption{\label{fig:FIG5} The SO(2,1) dynamics of one of the droplets in the circular quantum Hall droplet array (the one encircled by dashed lines in Fig.~\ref{fig:FIG4}). Figure (a) displays the temporal evolution of  $I(t)$ and $\boldsymbol L_{\rm p}(t)$, both calculated relative to center of the quantum Hall droplet. Figure (b) shows the snapshots of the droplet at various times. }
\end{figure}

In fact, it has been shown both theoretically and experimentally~\cite{Mukherjee2022} that a 1D quantum Hall droplet array can be formed by preparing the system in the stationary state $\psi(\br) = \tilde \psi_0(x)e^{-im\Omega xy}$ where $\tilde \psi_0(x)$ is determined by Eq.~(\ref{tildepsi}). Interestingly we find that the phase $\varphi = - m\Omega xy$ of the initial wave function, which gives rise to a sheared velocity profile,  always results in the formation of the 1D quantum Hall droplet array regardless of the initial density profile. Motivated by this insight, we find that it is possible to engineer the phase distribution of the initial wave function so as to achieve sheared velocity profiles conducive to the formation of 2D quantum Hall droplet arrays. For instance, a 2D  quantum Hall droplet array with the lattice configuration can be obtained by the following initial phase
\begin{align}
\varphi(x,y) = -m\Omega (x - 2jl_D) y 
\label{philattice}
\end{align}
for  $ (j-1)l_D/2\le x\le (j+1)l_D/2 $,  where $l_D$ is a parameter that determines the size of the eventual droplets and $j = 0, \pm 1, \pm 2,\cdots$. For the circular quantum Hall droplet array, the suitable initial phase (in polar coordinates) is
\begin{align}
\varphi(r,\theta) = m\Omega r_0^2 \theta,
\label{phiring}
\end{align}
for $0 < \theta \le 2\pi$ and $r_{\rm inner}\le r \le r_{\rm outer}$ , where $r_0 = (r_{\rm inner} + r_{\rm outer})/2$. The corresponding quantum Hall droplet arrays resulted from these two phase distributions are shown in Fig.~\ref{fig:FIG4}. As mentioned earlier, once the quantum Hall droplet arrays are formed, they are still in a dynamic state, as each of its droplets is in a constant, combined motion of stretching and rotation~\cite{SM}. To make contact with previously discussed single quantum Hall droplet dynamics, we calculate the moment of inertia and the physical angular momentum for a droplet in the circular array, which shows perfect agreements with Eqs.~(\ref{equ:I2})-(\ref{equ:LR2}) (see Fig.~\ref{fig:FIG5}). As a final example of the phase engineering approach to the formation of quantum Hall droplet arrays, we find that a 1D quantum Hall droplet array rotating at the cyclotron frequency  $\omega_c$ can be achieved by the initial phase
\begin{align}
\varphi(x,y) = m\Omega xy,
\label{phirotating}
\end{align}
which generates a velocity field $\bv = (2\Omega y, 0)$ as shown in Fig.~\ref{fig:FIG6}. The rotation of the array can be understood from Eq.~(\ref{phirtarrayr}) if we neglect the internal dynamics of individual quantum Hall droplets. 

\begin{figure}[t]
\centering
\includegraphics[width=0.48\textwidth]{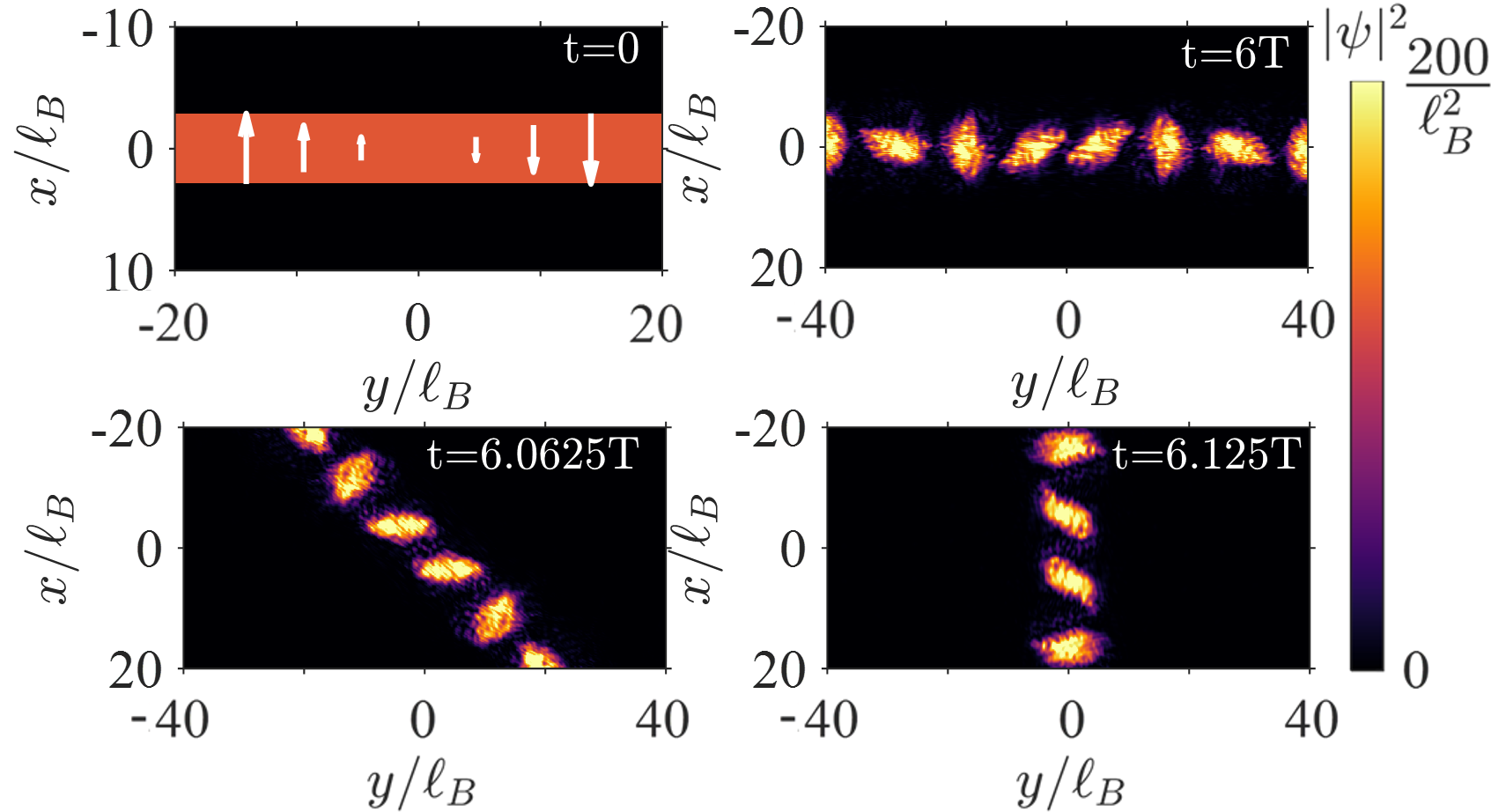} 
\caption{\label{fig:FIG6} A uniform strip of condensate with phase designed according to Eq.~(\ref{phirotating}) evolves into a quantum Hall droplet array rotating at the cyclotron frequency $\omega_c = 2\Omega$. Here the dimension of the initial condensate is $4\sqrt{2}\ell_B\times 80\sqrt{2}\ell_B$ and $N = 8\times 10^4$.}
\end{figure}

Bosonic quantum Hall droplets are very different from previously studied droplets in bosonic mixtures or in dipolar gases, which are stabilized by the balance between the mean-field attraction and the zero-point energy. Generally speaking, the bosonic quantum Hall droplets are stabilized by a balance of the outward mean-field pressure and an inward Coriolis force in the rotating frame. They are metastable states instead of the ground state of the system, the latter of which is in fact the vortex lattice. Therefore, the droplet's energy is not at a global minimum and it can not be obtained through a similar energetic analysis as the previous droplets. In this sense, their existence and stability are quite nontrivial.

Although our theory is strictly speaking only applicable to zero temperature systems, one can still obtain important insights about the influence of finite temperatures on the droplets. Through the Bogoliubov-de Gennes analysis, the droplet is dynamically stable against general excitations or fluctuations. Furthermore, our extensive simulations demonstrate that it is even stable when the fluctuation is quite large, due to nonlinear interaction. Therefore, one could foresee that the droplet should not be sensitive to the temperature whose main effect is thermal fluctuations and it may be stable at fairly large temperature regime  as long as the Bose condensation still occurs. This is also consistent with the experiment~\cite{Mukherjee2022}.

\emph {Conclusions.---}In summary, we have systematically studied the formation and dynamics of single quantum Hall droplet and droplet arrays in rapidly rotating BECs. Our theorem of localized many-body states of 2D interacting systems in a magnetic field provides a fundamental understanding for why quantum Hall droplet states exist in these rotating BECs. Our extensive numerical simulations provide further insight into the dynamical properties of the single quantum Hall droplet and droplet arrays. In particular, we have demonstrated the spontaneous formation of various quantum Hall droplet arrays out of phase-engineered initial states. Such an approach may be adopted in  experiments and pave the way for the future study of novel dynamics in rotating BEC.

{\it Acknowledgements.} We thank Yueheng Lan and Ping Fang for helpful discussions. This work is supported by National Key R\&D Program of China (Grant No. 2022YFA1404103), NSFC under Grant No. 12374243 (MYS) and No. 12474264 (ZGW), Xiaomi Young Talents Program, and Guangdong Provincial Quantum Science Strategic Initiative (Grant No. GDZX2404007).

\providecommand{\noopsort}[1]{}\providecommand{\singleletter}[1]{#1}%

\clearpage 
\onecolumngrid

\setcounter{equation}{0}
\setcounter{figure}{0}
\setcounter{table}{0}
\setcounter{page}{1}
\setcounter{section}{0}

\renewcommand{\theequation}{S\arabic{equation}}
\renewcommand{\thefigure}{S\arabic{figure}}
\renewcommand{\thetable}{S\arabic{table}}
\renewcommand{\thepage}{S\arabic{page}}
\renewcommand{\thesection}{\Roman{section}}


\begin{center}
    {\large\bfseries Supplementary Material for: \\
    Bosonic quantum Hall droplets in rapidly rotating  two-dimensional Bose-Einstein condensates} \\[.5cm]
    
\author{Zhen Cao}
\altaffiliation{These authors contributed equally to this work.}
\author{Siying Li}
\altaffiliation{These authors contributed equally to this work.}
\author{Zhendong Li}
\author{Xinyi Liu}
\affiliation{State Key Laboratory of Information Photonics and Optical Communications, School of Physical Science and Technology, Beijing University of Posts and Telecommunications, Beijing 100876, China.
}

\author{Zhigang Wu}
\email{wuzhigang@quantumsc.cn}
\affiliation{Quantum Science Center of Guangdong-Hong Kong-Macao Greater Bay Area (Guangdong), Shenzhen 508045, China
}
\author{Mingyuan Sun}
\email{mingyuansun@bupt.edu.cn}
\affiliation{State Key Laboratory of Information Photonics and Optical Communications, School of Physical Science and Technology, Beijing University of Posts and Telecommunications, Beijing 100876, China.
}

\date{\today}
\maketitle

\end{center}

%
%
%
%
%
%

This Supplemental Material includes the following three sections: (I) Derivation of Eq.~(4) of the main text (II) Bogoliubov-de Gennes analysis of the single quantum Hall droplet dynamics and (III) Dynamics of the quantum Hall droplet arrays.  
\section{Derivation of Eq.(4) in the main text }
In this section, we derive Eq.(4) in the main text. Here, we consider the many-body Hamiltonian of a 2D interacting system in a magnetic field which reads
\begin{align}
\hat H = \sum_j  \frac{1}{2m} \left [\hat \bp_j - \bA(\hat\br_j)\right ]^2   + \sum_{j<k} V(\hat \br_j - \hat \br_k),
\end{align}
where $\bA(\br) = m\Omega\hat z \times \br$ is the vector potential of a uniform magnetic field $\boldsymbol B = 2m\Omega\hat z$ and  $V(\br)$ is the pairwise particle interaction.  
We are interested in the dynamics of the center of mass operator
\begin{align}
\hat \bR = \frac{1}{N} \sum_j \hat \br_j
\end{align}
and the total canonical momentum operator
\begin{align}
\hat \bP = \sum_j \hat \bp_j.
\end{align}
The key observation is that these degrees of freedom are decoupled from the internal degrees of freedom. Indeed,  the Heisenberg equation of motion for these operators are in the following closed form
\begin{align}
\frac{d \hat \bR}{dt} &= \frac{1}{M}\hat \bP - \Omega \hat \bz \times \hat \bR \\
\frac{d\hat \bP}{dt}& = -M\Omega^2 \hat \bR - \Omega \hat \bz \times \hat\bP,
\end{align}
where $M = N m$ and $\hat \bz$ is the unit vector along the $z$-direction. The equation for the momentum operator can be written as
\begin{align}
\frac{d\hat \bP}{dt} & = -M\Omega \left (\frac{1}{M} \hat \bz \times \hat \bP + \Omega \hat \bR \right )  \nonumber \\
& = -M\Omega \hat\bz \times \left (\frac{1}{M}  \hat \bP - \Omega \hat \bz \times\hat \bR \right ) \nonumber \\
& = -M\Omega \hat\bz \times \frac{d\hat \bR}{dt}.
\label{Peqspec}
\end{align}
Thus we find 
\begin{align}
\ddot{\hat \bR} &= \frac{1}{M} \dot{\hat \bP} -\Omega \hat \bz \times \dot {\hat \bR} \nn \\
& = -2\Omega \hat\bz \times \dot{\hat \bR}
\end{align}
and 
\begin{align}
\dddot{\hat\bR} &= -2\Omega \hat \bz \times \ddot{\hat \bR} \nn \\
& = 4\Omega^2 \hat\bz \times (\hat \bz \times \dot{\hat \bR}) \nn \\
& = -4\Omega^2 \dot{\hat \bR}.
\end{align}
The above equation can be solved as
\begin{align}
\dot{\hat \bR} (t) &= \dot{\hat \bR} (0) \cos 2\Omega t + \frac{\ddot{\hat \bR}(0)}{2\Omega} \sin 2\Omega t \nn \\
& = \dot{\hat \bR} (0) \cos 2\Omega t - \hat\bz \times \dot{\hat \bR}(0)\sin 2\Omega t.
\end{align}
More explicitly we have
\begin{align}
\dot{\hat R}_x(t) &= \dot{\hat R}_x(0)\cos 2\Omega t  +  \dot{\hat R}_y(0) \sin 2\Omega t  \nn \\
\dot{\hat R}_y(t) &= -\dot{\hat R}_x(0)\sin 2\Omega t  +  \dot{\hat R}_y(0) \cos 2\Omega t.
\end{align}
Integrating these equations we obtain 
\begin{align}
\hat R_{x}(t) &= \hat R_{x}(0)  + \frac{\dot{\hat R}_y(0)}{2\Omega} + \frac{\dot{\hat R}_x(0)}{2\Omega}\sin 2\Omega t -    \frac{\dot{\hat R}_y(0)}{2\Omega} \cos 2\Omega t \\ 
\hat R_{y}(t) &= \hat R_{y}(0)  - \frac{\dot{\hat R}_x(0)}{2\Omega} + \frac{\dot{\hat R}_x(0)}{2\Omega}\cos 2\Omega t +    \frac{\dot{\hat R}_y(0)}{2\Omega} \sin 2\Omega t .
\end{align}
Expressing $\dot{\hat \bR}(0)$ in terms of $\hat \bR (0)$ and $\hat \bP (0)$, we have 
\begin{align}
\label{Rxt}
\hat R_{x}(t) &= \frac {1+\cos 2\Omega t }{2}{\hat R_{x}(0)}  +  \frac{\sin 2\Omega t }{2} \hat R_{y}(0)  + \frac{ \sin 2\Omega t  }{2M\Omega}  \hat P_x(0) +\frac{ 1-\cos 2\Omega t  }{2M\Omega} \hat P_{y} (0) \\
\hat R_{y}(t) &= -  \frac{\sin 2\Omega t }{2} \hat R_{x}(0)  + \frac {1+\cos 2\Omega t }{2}{\hat R_{y}(0)}   - \frac{ 1 - \cos 2\Omega t  }{2M\Omega}  \hat P_x(0) +\frac{ \sin 2\Omega t  }{2M\Omega} \hat P_{y} (0).
\label{Ryt}
\end{align}
Integrating Eq.~(\ref{Peqspec}) we find
\begin{align}
\label{Pxt}
\hat P_x(t) - \hat P_x(0) &= M\Omega \left [ \hat R_{y}(t) - \hat R_{y}(0)\right ] \\
\hat P_{y}(t) - \hat P_{y}(0) &= -M\Omega \left [ \hat R_{x}(t) - \hat R_{x}(0)\right ].
\label{Pyt}
\end{align}
More specifically, we have
\begin{align}
\hat P_x(t)  & =-  \frac{M\Omega\sin 2\Omega t }{2} \hat R_{x}(0)  - \frac {M\Omega(1-\cos 2\Omega t) }{2}{\hat R_{y}(0)}   + \frac{ 1 + \cos 2\Omega t  }{2}  \hat P_x(0) +\frac{ \sin 2\Omega t  }{2} \hat P_{y} (0)   \\
\hat P_{y}(t)  & = \frac {M\Omega(1-\cos 2\Omega t )}{2}{\hat R_{x}(0)}  -  \frac{M\Omega\sin 2\Omega t }{2} \hat R_{y}(0)  - \frac{ \sin 2\Omega t  }{2}  \hat P_x(0) +\frac{ 1+\cos 2\Omega t  }{2} \hat P_{y} (0)  .
\end{align}
Using Eqs.~(\ref{Rxt})-(\ref{Ryt}) and Eqs.~(\ref{Pxt})-(\ref{Pyt}), we find  Eq.(4) of the main text, i.e., 
\begin{align}
    \bp_0\cdot \hat \bR(-t) - \br_0\cdot \hat \bP(-t) = \bp_0(t)\cdot \hat \bR(0) - \br_0(t)\cdot \hat \bP(0) 
\end{align}
where 
\begin{align}
p_{0,x}(t) & = \frac {1+\cos 2\Omega t }{2}p_{0,x} +  \frac{\sin 2\Omega t }{2} p_{0,y} -\frac{M\Omega\sin 2\Omega t }{2}  r_{0,x} - \frac {M\Omega(1-\cos 2\Omega t )}{2}  r_{0,y} \\
p_{0,y}(t) & =  -\frac{\sin 2\Omega t }{2} p_{0,x} + \frac {1+\cos 2\Omega t }{2}p_{0,y} +  \frac {M\Omega(1-\cos 2\Omega t )}{2} r_{0,x}  - \frac{M\Omega\sin 2\Omega t }{2}  r_{0,y} \\
r_{0,x}(t) & = \frac {\sin 2\Omega t }{2M\Omega}p_{0,x} +  \frac {1-\cos 2\Omega t }{2M\Omega} p_{0,y} +\frac{1 + \cos 2\Omega t }{2}  r_{0,x} + \frac {\sin 2\Omega t }{2}  r_{0,y} \\
r_{0,y}(t) & = - \frac{1- \cos 2\Omega t }{2M\Omega} p_{0,x} +  \frac {\sin 2\Omega t }{2M\Omega}p_{0,y} -  \frac {\sin 2\Omega t }{2} r_{0,x} + \frac{1+\cos 2\Omega t }{2}  r_{0,y}. \\
\end{align}
It is not difficult to see that we can write $\br_0(t)$ and $\bp_0(t)$ in the following compact forms
\begin{align}
\br_0(t) &= \br_0 - \bd(0) +\bd(t) \\
\bp_0(t)& = M\dot{\br}_0(t) + M\Omega \hat z \times \br_0(t),
\end{align}
where $\br_0(0) = \br_0$, $\bp_0(0) = \bp_0$ and  $\bd(t) = (d_x(t),d_y(t))$ is given by
\begin{align}
 \begin{pmatrix}
 d_x(t) \\ d_y(t)
 \end{pmatrix} =  \begin{pmatrix}
 \cos\omega_c t & \sin\omega_c t \\
 -\sin\omega_c t& \cos\omega_c t
 \end{pmatrix}  \begin{pmatrix}
 d_x(0) \\d_y(0)
 \end{pmatrix},
 \end{align} 
 with $\bd (0) = \frac{1}{2}\br_0 + \frac{1}{2M\Omega}\hat z\times\bp_0$ and $\omega_c = 2\Omega$.

\section{Bogoliubov-de Gennes analysis of the single quantum Hall droplet dynamics}
In this section, we provide a more detailed analysis of the linear and non-linear stability of the single quantum Hall droplet. Based on Eqs.(18)-(19) of the main text, one can obtain the excitation spectrum and eigen-modes of the stationary quantum Hall droplet. Then, the dynamics of the droplet under perturbation can be investigated in the linear regime by expanding the initial condensate wave function over the eigen-modes as follows
\begin{align}
\psi(\br) & \equiv  \Phi_0(\br) + \delta \psi(\br) \nn \\
&= \Phi_0(\br) + \sum_{nl} \left [ c_{nl}  u_{nl} (\br) -c^*_{nl}   v^*_{nl}(\br)  \right ],
\end{align}
\begin{figure}[h]
\centering
\includegraphics[width=0.95\textwidth]{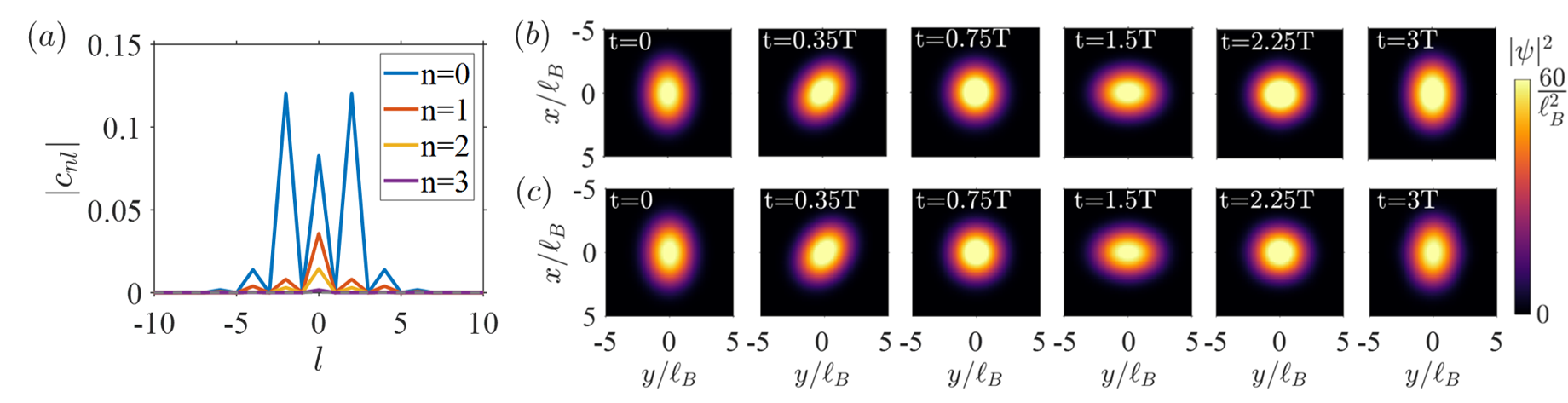}
\caption{\label{fig:FIGS1} Dynamic evolution of a quantum Hall droplet perturbed from equilibrium in the linear regime. (a) the expansion coefficient of the initial wavefunction over the BdG eigen-modes, obtained from Eq.~\ref{cnl}; (b) the calculated droplet dynamics through Eq.~\ref{BdGdynamics} with BdG analysis; (c) the calculated droplet dynamics by solving the GP equation. Here
$T = 2\pi/\Omega$, $N = 800$ and $\epsilon=0.4$ for the initial condensate. }
\end{figure}
\begin{figure}[h]
\centering
\includegraphics[width=0.95\textwidth]{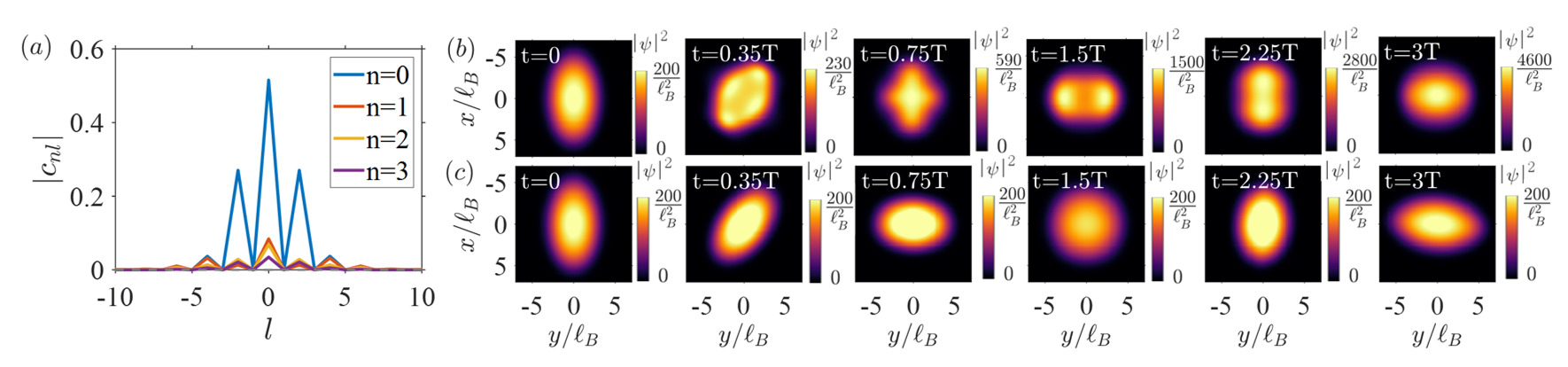}
\caption{\label{fig:FIGS2} Dynamic evolution of a quantum Hall droplet perturbed from equilibrium in the nonlinear regime. (a) the expansion coefficient of the initial wavefunction over the BdG eigen-modes, obtained from Eq.~\ref{cnl}; (b) the calculated droplet dynamics through Eq.~\ref{BdGdynamics} with BdG analysis; (c) the calculated droplet dynamics by solving the GP equation. Here
$T = 2\pi/\Omega$, $N = 8000$ and $\epsilon=0.5$ for the initial condensate. }
\end{figure}
where $c_{nl}$ are the coefficients to be determined. Using the orthonormal relations of the Bogoliubov amplitudes 
\begin{align}
\int d\br \left [  u_{nl}(\br)u^*_{n'l'}(\br)  -  v_{nl}(\br) v^*_{n'l'}(\br)\right ] &= \delta_{nn'}\delta_{ll'} \\
\int d\br \left [  u_{nl}(\br)v_{n'l'}(\br)  -  v_{nl}(\br) u_{n'l'}(\br)\right ] &= 0,
\end{align}
we find
\begin{align}
c_{nl}  = \int d\br \left [  u_{nl}^*(\br) \delta \psi(\br) + v_{nl}^*(\br) \delta \psi^*(\br) \right ].
 \label{cnl}
\end{align}
Within the Bogoliubov-de Gennes (BdG) framework, the time evolution of the wavefunction can be expressed as
\begin{align}
 \psi(\br,t)  = \psi_0(\br)e^{-i\mu t } +e^{-i\mu t}\sum_{nl} \left [ c_{nl} u_{nl} (\br) e^{-i\epsilon_{nl} t} - c^*_{nl} v^*_{nl}(\br) e^{i\epsilon_{nl} t} \right ].
 \label{BdGdynamics}
\end{align}
A typical perturbation can be imposed by squeezing the condensate along one direction using the following anisotropic trapping potential 
\begin{align}
 V_{\rm tr} (\br) = \frac{1}{2} m\Omega^2 [(1-\epsilon) x^2 + (1+\epsilon)y^2].
 \end{align} 
 
Preparing the condensate at $t=0$ in the above trapping potential, we perform extensive computations of the quantum Hall droplet dynamics in various parameter regions both by using Eq.~(\ref{BdGdynamics}) and by solving the time-dependent Gross-Pitaevskii equation (see Eq.(10) of the main text). We find that the two methods agree with each other in the linear regime as expected (see the example displayed in Fig.~\ref{fig:FIGS1}). These results demonstrate the stability of the quantum Hall droplet around the stationary state. Surprisingly, even in the highly nonlinear regime, where the BdG analysis [i.e., Eq.~(\ref{BdGdynamics})] is not applicable, we find the quantum Hall droplet is still stable, as shown in Fig.~\ref{fig:FIGS2}.

\section{Dynamics of the quantum Hall droplet arrays}
\begin{figure}[t]
\centering
\includegraphics[width=0.75\textwidth]{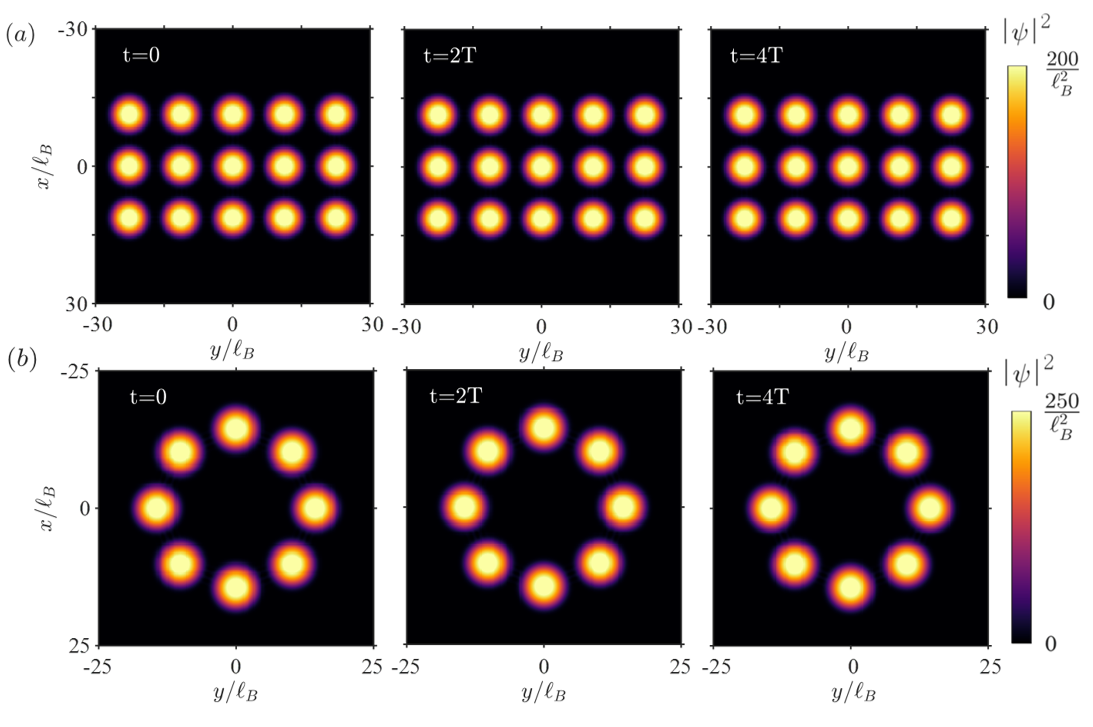}
\caption{\label{fig:FIGS3} The  stationary quantum Hall  droplet arrays at different times. The parameters here are $N = 10^5$ (a) and $N = 8\times 10^4$ (b).}
\end{figure}
\begin{figure}[h]
\centering
\includegraphics[width=0.99\textwidth]{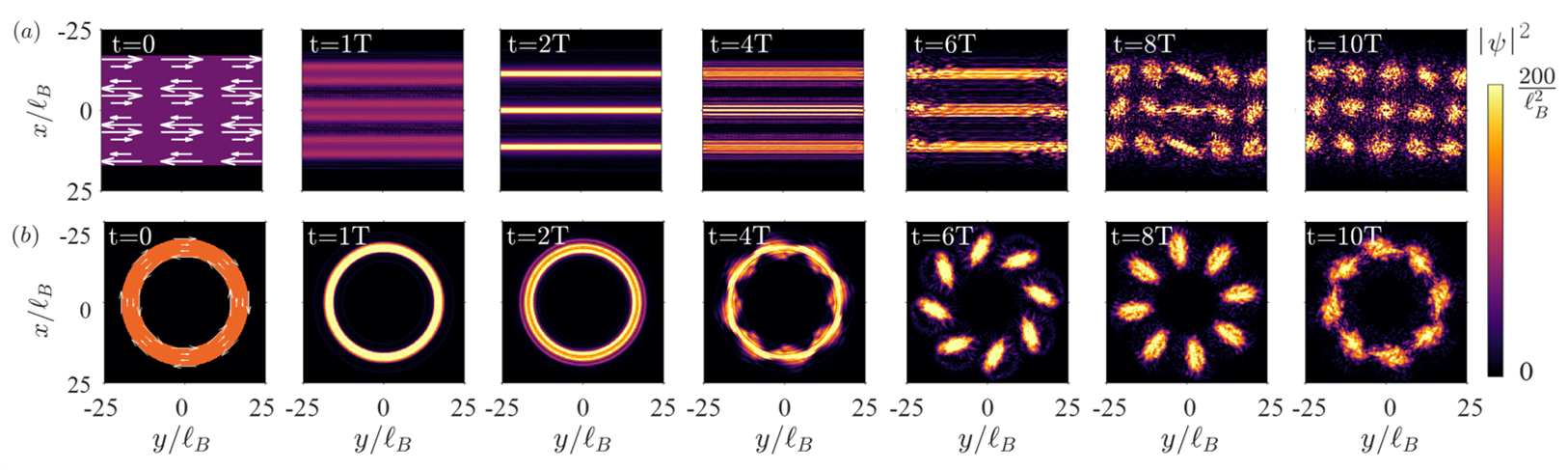}
\caption{\label{fig:FIGS4} The formation process of dynamic quantum Hall droplet arrays from a uniform condensate. (a) The phase is designed according to Eq.~(25) of the main text, which gradually evolves into a droplet lattice.  Here $N = 2.4\times10^5$ and the dimension  of the initial condensate is $24\sqrt{2}\ell_B\times 80\sqrt{2}\ell_B$. (b) A ring shaped uniform condensate with phase designed according to Eq.~(26) of the main text evolves into a circular droplet array. Here $N = 8\times10^4$ and the  outer radius and the width of initial condensate are  $14\sqrt{2}\ell_B$ and $4\sqrt{2}\ell_B$ respectively.}
\end{figure}
\begin{figure}[h]
\centering
\includegraphics[width=0.99\textwidth]{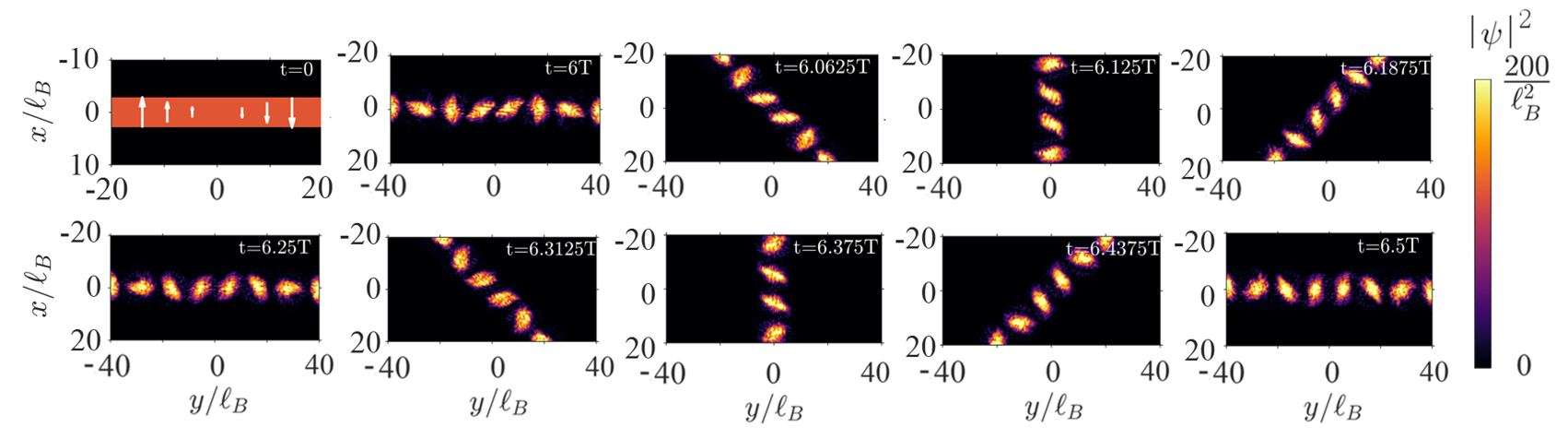}
\caption{\label{fig:FIGS5} A uniform strip of condensate with phase designed according to Eq.~(27) of the main text evolves into a droplet array rotating at the cyclotron frequency $\omega_c = 2\Omega$. Here the dimension of the initial condensate is $4\sqrt{2}\ell_B\times 80\sqrt{2}\ell_B$ and $N = 8\times 10^4$. }
\end{figure}

In this section, we provide more details on the dynamic evolution of the quantum Hall droplet arrays. In Fig.~3  of the main text, we have only displayed the stationary droplet arrays constructed using Eq.~(22) of the main text at a specific time $t = 4T$. Here in Fig.~\ref{fig:FIGS3} we provide snapshots of the arrays at $t = 0, 2T, 4T$, demonstrating the stationary character of these droplet arrays. Furthermore, in Fig.~\ref{fig:FIGS4} we show a more detailed evolution process of the dynamic formation of the droplet arrays corresponding to those in Fig.~4 of the main text.  Lastly, in Fig.~\ref{fig:FIGS5}   we display the evolution of the rotating 1D quantum Hall droplet array (Fig.~(6) of the main text) within a full period of $T/2$ to show that once the droplet array is formed it indeed performs the rigid-body-like rotation at the cyclotron frequency $\omega_c = 2\Omega$. 
\hfill\hfill


\end{document}